\documentclass[aps,prl,twocolumn,showpacs,floatfix,superscriptaddress]{revtex4}
\usepackage{graphicx,amsmath}

\begin{document}

\title{Extracting Multidimensional Phase Space Topology from Periodic Orbits}
\author{Stephan Gekle}
\author{J\"org Main}
\affiliation{Institut f\"ur Theoretische Physik 1, Universit\"at Stuttgart,
 70550 Stuttgart, Germany}
\author{Thomas Bartsch}
\author{T. Uzer}
\affiliation{Center for Nonlinear Science, Georgia Institute of Technology, 
 Atlanta, Georgia 30332-0430, USA}
\date{\today}

\begin{abstract}
We establish a hierarchical ordering of periodic orbits in a strongly
coupled multidimensional Hamiltonian system. Phase space structures can
be reconstructed quantitatively from the knowledge of periodic orbits
alone. We illustrate our findings for the hydrogen atom in crossed electric
and magnetic fields.
\end{abstract}


\maketitle

Periodic motion is widely recognized as the most prominent feature in a
wide range of dynamical systems, from astronomy \cite{Copenhagen, Jorba_2,
Poincare_cite}, molecular vibrations \cite{Chemistry_1, Joyeux97},
chemical reactions \cite{Chemistry_3}, particle accelerators
\cite{Resonance_1}, atomic and molecular physics \cite{Gutzwiller_book,
Quasi_Landau, Wang_planar} to fluid dynamics, e.g., statistics of turbulent
flow \cite{Fluids_1}. Because periodic orbits are the fundamental building
blocks of the dynamics, they offer a road map to the intricate geometrical
and dynamical structure in a multidimensional phase space. Unfortunately,
this map is hard to decipher: High-dimensional systems usually possess
enormous numbers of periodic orbits whose geometric appearance in
configuration space gives no useful hint at their systematics
\cite{Quasi_Landau,Wang_planar}. Indeed, it is not even clear \emph{a
priori} that a systematic organization of periodic orbits should exist at
all.

\begin{figure}[b]
  \includegraphics[width=.95\columnwidth]{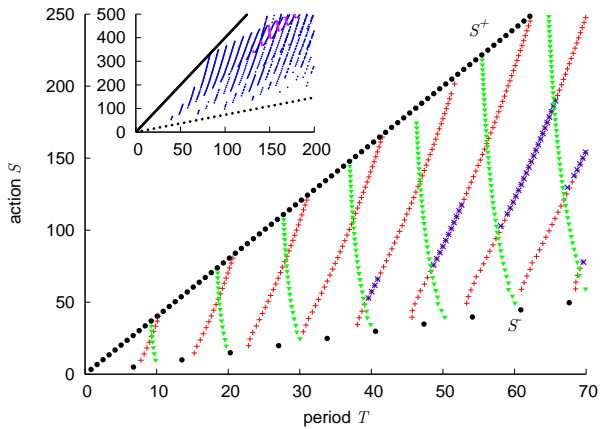}
  \caption{(color online).
  Periodic orbits at $E=-1.4\,\text{a.u.}$, $F=0.5\,\text{a.u.}$
  FPOs and their
  repetitions are shown with black circles, $T_2^\text{p}$ POs with
  red plus symbols, $T_2^\text{n}$ with green triangles,
  $T_3^\text{p}$ with blue crosses, and $T_3^\text{n}$ with magenta diamonds.
  The inset presents the 3-torus POs in an enlarged $T$-$S$ range.}
\label{fig:TS}
\end{figure}

The hydrogen atom in crossed electric and magnetic fields is typical in
this respect. It serves as a paradigm of strongly coupled multidimensional
systems because it can be investigated both experimentally and
theoretically \cite{Quasi_Landau, Ericson_fluctuations_2,
Wang_planar}. However, after two decades of intense scrutiny, the overall
phase space structure still defies a complete understanding. In particular,
no pertinent symbolic dynamics are known. Although extensive lists of
periodic orbits (POs) are available in the literature
\cite{Floethmann_Kepler,Floethmann_Moon,Quasi_Landau,Wang_planar}, no
comprehensive ordering scheme for POs in three degrees of freedom has been
proposed so far, nor has it been shown how POs can provide insight into
higher-dimensional phase space structures. That such a scheme should exist
becomes clear from Fig.~\ref{fig:TS}: We find that the periods and actions
of most periodic orbits (POs) in the crossed-fields hydrogen atom fall into
clearly discernible series.  It is the central aim of this paper to
elucidate the dynamical origin of this remarkable structure. We will show
that a hierarchical ordering of periodic orbits arises from the hierarchy
of invariant tori in an integrable limiting case, and we will develop the
tools needed to characterize this organization
in a non-integrable system beyond two degrees of freedom. In the course of
this work, the nomenclature used in Fig.~\ref{fig:TS} will be made clear.

The Hamiltonian for the electronic motion in crossed-fields hydrogen reads,
in atomic units,
\begin{equation}
 H = \frac{1}{2}\mathbf{p}^2 - \frac{1}{r} + \frac{B}{2}\left(p_yx-p_xy\right)
 + \frac{B^2}{8}\left(x^2+y^2\right) - Fx.
\label{eqn:Hamiltonian}
\end{equation}
Here $\mathbf r=(x,y,z)$ are the usual Cartesian coordinates, $\mathbf
p=(p_x,p_y,p_z)$ the conjugate momenta, and $r=\sqrt{x^2+y^2+z^2}$.  The
magnetic field $B$ points in the $z$ direction and the electric field $F$ in
the $x$ direction. Exploiting the scaling property of the classical
Hamiltonian (\ref{eqn:Hamiltonian}), we can set $B=1\,\text{a.u.}$ without
loss of generality. We take $F=0.5\,\text{a.u.}$ and consider two energies,
$E=-1.5\,\text{a.u.}$ and $E=-1.4\,\text{a.u.}$, slightly below and slightly
above the classical ionization threshold $E_{\mathrm{I}}=-2\sqrt{F}$.  To
identify POs, we employ a simple shooting algorithm, which like any other
numerical PO search is not guaranteed to find all POs. However, our method of
torus construction does not require the knowledge of the complete set of POs.

\begin{figure}
  \includegraphics[width=.95\columnwidth]{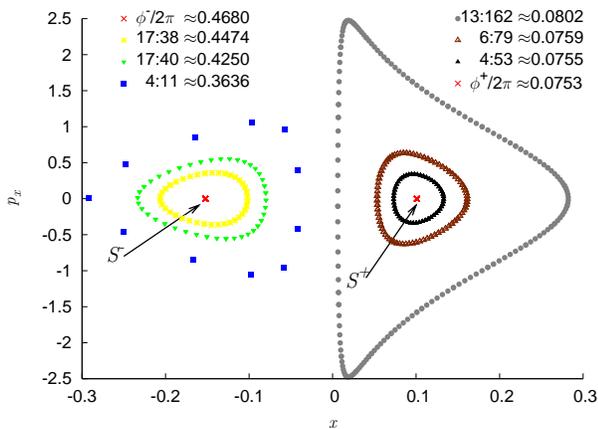}
  \caption{(color online).
  Periodic orbits in the Poincar\'e surface of section $y=0$
  for the planar subsystem, $E=-1.5\,\text{a.u.}$, $F=0.5\,\text{a.u.}$
  Non-fundamental POs are labeled by their winding ratios.}
\label{fig:sos}
\end{figure}

To illustrate the mechanism that leads to the hierarchical ordering of POs,
we will first study a two-dimensional subsystem of the crossed-fields
hydrogen atom, namely the $x$-$y$-plane perpendicular to the magnetic
field.  Its dynamics can conveniently be discussed with the help of a
Poincar\'e surface of section such as Fig.~\ref{fig:sos}. The plot shows
two elliptic (stable) fundamental periodic orbits (FPO), called $S^+$ and
$S^-$, surrounded by POs of larger periods.  The latter arise in
stable/unstable pairs according to the Poincar\'e-Birkhoff theorem
\cite{Arnold}. (For graphical reasons, only one orbit of each pair is
displayed in Fig.~\ref{fig:sos}.) They can be thought of as being generated
from the breakup of a resonant invariant torus in an integrable limiting
case of the dynamics. In the situation of Fig.~\ref{fig:sos}, we say that
the FPOs serve as the organizing centers for the longer orbits that
surround them.

Similar structures can be expected to exist in higher dimensions and have
indeed been observed in integrable systems \cite{Joyeux97}. Given that in a
Hamiltonian system of $L$ degrees of freedom there can be invariant tori of
dimension up to $L$, one might expect to find an entire hierarchy: FPOs serve
as organizing centers for families of POs that arise from the breakup of
2-dimensional tori (2-torus POs), which in turn organize families that are
generated from 3-dimensional tori (3-torus POs), and so on up to the maximum
dimension.  However, Poincar\'e surface of section plots are hard to visualize
beyond two degrees of freedom. Therefore, they do not provide a practical tool
to identify these families of POs and to diagnose their mutual
relationships. We will here introduce a quantitative method to accomplish
these two tasks in a multidimensional system and we will demonstrate in a
3-dimensional example that the hierarchy outlined above does indeed exist. The
hierarchy of POs that we derive mirrors the hierarchy and topology of their
originating invariant tori in the integrable limit, which is also the
hierarchy of surviving invariant Kolmogorov-Arnold-Moser (KAM) tori
\cite{Arnold} of the full non-integrable dynamics.

The key conceptual tools that we use to elucidate this hierarchy are
action variables and winding numbers.  In the integrable limit, any motion
is confined to an invariant torus \cite{Arnold} that is characterized by a
set of conserved action variables $\mathbf{I}$. Their conjugate angles
$\boldsymbol{\theta}$ determine the position on the torus. If an $n$-torus
carries POs, integer winding number $w_1,\dots,w_n$ can be assigned to it
so that the angle $\theta_i$ runs from 0 to $2\pi\cdot w_i$ before the
orbit closes. Because winding numbers are topological properties, they
remain meaningful in the non-integrable case, where the resonant tori have
broken up into isolated POs.  Using the winding numbers $\mathbf{w}$,
the total action can be written as
$S=\mathbf{w}\cdot\mathbf{I}(\mathbf{w})$, where $\mathbf{I}(\mathbf{w})$
are the action coordinates of the originating torus. In fact, the actions
depend only on the ratios of winding numbers, e.g., $w_1/w_2$ and $w_3/w_2$
for 3-torus POs or $w_1/w_2$ for 2-torus POs.

While the assignment of winding numbers to a given PO is simple
with the help of a Poincar\'e surface of section plot, it is in itself a
difficult problem in more than two degrees of freedom. 
The geometric appearance of
the orbits in configuration space does not provide a useful guide. A major
part of this Letter will therefore be devoted to the
development of a viable method to assign winding numbers to individual POs.

\begin{figure}
  \includegraphics[width=.95\columnwidth]{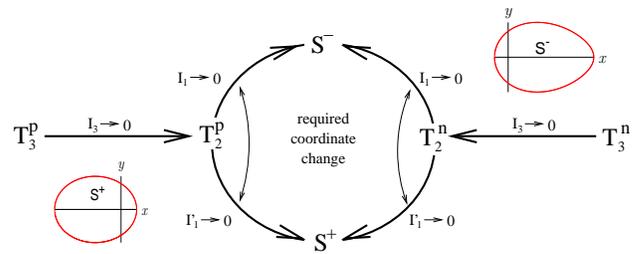}
  \caption{(color online).
   The hierarchy of $n$-torus POs in the crossed-fields hydrogen atom at
   $E=-1.5\,\text{a.u.}$, $F=0.5\,\text{a.u.}$
   Insets show the FPOs in configuration space.}
\label{fig:genealogy}
\end{figure}

Once winding numbers (and the associated action variables, the calculation
of which is then straightforward) are available, the situation in
Fig.~\ref{fig:sos} can be characterized by the following quantitative
criteria: 
(i) The stability angles $\phi_1^\pm$ of the FPOs $S^\pm$ 
(i.e.~the phase angles of the unimodular eigenvalues of their stability 
matrices) describe the rotation that each FPO imposes upon its neighborhood.
The winding ratios of the 2-torus POs converge toward $\phi_1^\pm/2\pi$ as 
the FPOs are approached. 
(ii) In the same limit, the action variable corresponding to the degree 
of freedom along the FPO converges to the action of the FPO, and
(iii) the action variable for the motion transverse to the FPO, which is
given by the area the (original) invariant torus encloses in the Poincar\'e
plane, tends to zero. 
We will use these three criteria to identify
analogous situations in higher dimensions, and we will show that in a
similar manner families of 2-torus POs can themselves serve as organizing
centers for families of 3-torus POs. Notice that although the choice of
action-angle coordinates and the associated winding numbers is not unique,
within the hierarchical structure the organizing center imposes a specific
coordinate system (along/transverse to the FPO) upon the family it
organizes.

POs that are remnants of the same torus have nearly identical periods and
actions, thus each point in Fig.~\ref{fig:TS} represents POs generated from
a single torus. As in the two-dimensional case, 2-torus POs occur in
doublets. By contrast, 3-torus POs arise in groups of four \cite{Meiss}. At
the basis of the hierarchy of POs in the crossed-fields hydrogen atom at
$E=-1.5\,\text{a.u.}$ there are the FPOs (or 1-torus POs) $S^+$ and $S^-$
that were described in \cite{Floethmann_Kepler,Floethmann_Moon}. They are
shaped nearly like Keplerian ellipses in the invariant $x$-$y$-plane.
2-torus and 3-torus POs are arranged in series with positive or negative
slopes in the $T$-$S$-diagram of Fig.~\ref{fig:TS}, which form the
$T^\text{p}_{2,3}$ and $T^\text{n}_{2,3}$ families, respectively. The POs
contained in the $x$-$y$ plane and illustrated in Fig.~\ref{fig:sos} form
the family $T^\text{p}_2$. The series of 2-torus POs end at boundary lines
marked by the FPOs and their repetitions. This observation suggests that
each of the FPOs serves as an organizing center for both families
$T_2^\text{p}$ and $T_2^\text{n}$. We will show below that this is indeed
the case. (Notice that in three degrees of freedom each FPO has two
transverse degrees of freedom characterized by two different stability
angles. It can therefore organize two different families of 2-torus POs.)
In addition, we will see that the families $T_2^\text{p,n}$ themselves act
as organizing centers for the families $T_3^\text{p,n}$ of 3-torus POs. The
entire hierarchy thus obtained is illustrated schematically in
Fig.~\ref{fig:genealogy}. At the energy $E=-1.4\,\text{a.u.}$, above the
ionization saddle point, we find the same general structure, except that
the FPO $S^-$ is surrounded by a region of ionizing trajectories and thus
cannot serve as an organizing center for the $T_2^\text{p,n}$.

Winding numbers for the 2-torus POs $T^\text{p}_2$ and $T_2^\text{n}$ can
be assigned as follows: Counting the series in Fig.~\ref{fig:TS} yields the
first winding number $w_1$. The second winding number $w_2$ is found by
counting the POs within one series from bottom to top. It remains only to
determine with what values the counting is to start in both cases. To this
end, we use Fourier expansions of the coordinate functions $x(t)$ and $z(t)$.
These spectra show two major peaks.
One of them agrees with the series number $w_1$ (starting with
$w_1=1$ in the leftmost series), whereas the location of the second is
identified with the second winding number $w_2$ (starting with $w_2=2$ for
the shortest orbit in the leftmost series). The Fourier expansions thus
confirm the validity of the simple counting scheme described above.

\begin{figure}
  \includegraphics[width=.94\columnwidth]{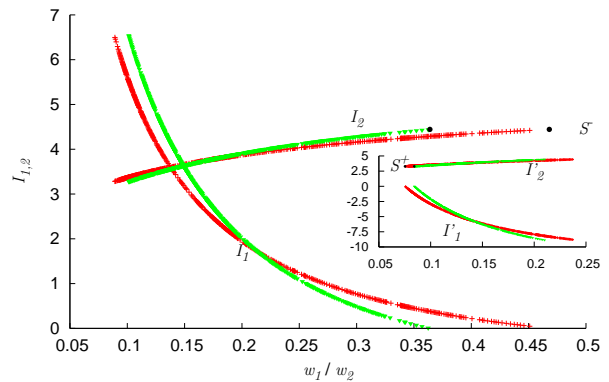}
  \caption{(color online).
  Action variables for the $T_2^\text{p}$ (red plus symbols) and
  the $T_2^\text{n}$ (green triangles) families at $E=-1.5\,\text{a.u.}$,
  $F=0.5\,\text{a.u.}$ Black dots show the action and the stability angles
  of the organizing center $S^-$.
  The inset presents the same situation in a different action-angle
  coordinate system, in which $S^+$ can be identified as another organizing
  center of the $T_2^\text{p,n}$ families.}
\label{fig:2_tori_actions}
\end{figure}

For the 2-torus POs we obtain the action coordinates $\mathbf{I}$ shown in
Fig.~\ref{fig:2_tori_actions} together with the actions and stability
angles $\phi_{1,2}$ of the FPOs $S^\pm$.  The limiting values for high
winding ratios of the $T_2^\text{p}$ and $T_2^\text{n}$ families coincide
with $\phi^-_1/2\pi$ and $\phi^-_2/2\pi$ of $S^-$, respectively. As the maximum
winding ratio is approached, $I_2$ converges towards the action of the FPO,
whereas $I_1$ vanishes. According to the three criteria listed above, we
can therefore conclude that the FPO $S^-$ serves as an organizing center
for the two families $T_2^\text{p}$ and $T_2^\text{n}$ of 2-torus POs, and
we can identify the angles conjugate to $I_2$ and $I_1$ as the degrees of
freedom along and transverse to the central FPO $S^-$. 

At the lower limits of the winding ratios in Fig.~\ref{fig:2_tori_actions},
none of the action variables converges to zero, and it seems, therefore,
that, contrary to the hypothesis derived from Fig.~\ref{fig:TS}, the
$T_2^\text{p}$ and $T_2^\text{n}$ families are not organized by $S^+$ the
way they are by $S^-$. However, this difference is an artifact of the
action-angle coordinate system. In fact, Fourier spectra suggest that for
2-torus POs close to $S^+$ a set of winding numbers $\mathbf{w'}$ should be
used that starts the counting scheme with $w_2=4$ for the shortest orbit in
the leftmost series. This set of winding numbers leads to the action
variables $\mathbf{I'}$ used in
the inset of Fig.~\ref{fig:2_tori_actions}.  In this coordinate system,  the
approach to $S^+$ satisfies the three criteria set up above, whereas the
approach to $S^-$ does not. Therefore, both FPOs $S^+$ and
$S^-$ serve as organizing centers for the two families of 2-tori POs in an
equal manner. (The POs in Fig.~\ref{fig:TS} are thus labeled in the
coordinate system induced by the nearest FPO.)

For the $T_2^\text{p}$, the conclusion that $S^+$ and $S^-$ serve as
organizing centers for families of 2-torus POs has already been reached
from the Poincar\'e surface of section plot in Fig.~\ref{fig:sos}.  Notice
that in spite of its intuitive appeal the Poincar\'e plot is somewhat
misleading because it seems to show two distinct families of 2-torus
POs. In fact, however, the $T$-$S$-plot of Fig.~\ref{fig:TS} indicates, and
the assignment of winding numbers confirms, that all planar 2-torus POs
form the single family $T_2^\text{p}$.
A two-dimensional Poincar\'e surface of section is entirely unsuitable to
analyze the geometry of the $T_2^\text{n}$ because these POs do not lie in
a two-dimensional plane. Nevertheless, the quantitative method developed
here demonstrates that the relation of $T_2^\text{n}$ to the FPOs $S^\pm$
is the same as that of the $T_2^\text{p}$.
Apart from avoiding this ambiguity, our method for the analysis of phase
space topology is also more general than Poincar\'e surface of section
techniques in that it can now readily be applied to the 3-torus POs.

The 3-torus POs in Fig.~\ref{fig:TS} fall into the same series as the
2-torus POs, but they possess a third winding number $w_3$. The latter
manifests itself in an additional major peak in the Fourier spectra
of the coordinates,
and therefore can be assigned by a straightforward extension of the 
technique used to classify the 2-torus POs.

\begin{figure}
  \includegraphics[width=.93\columnwidth]{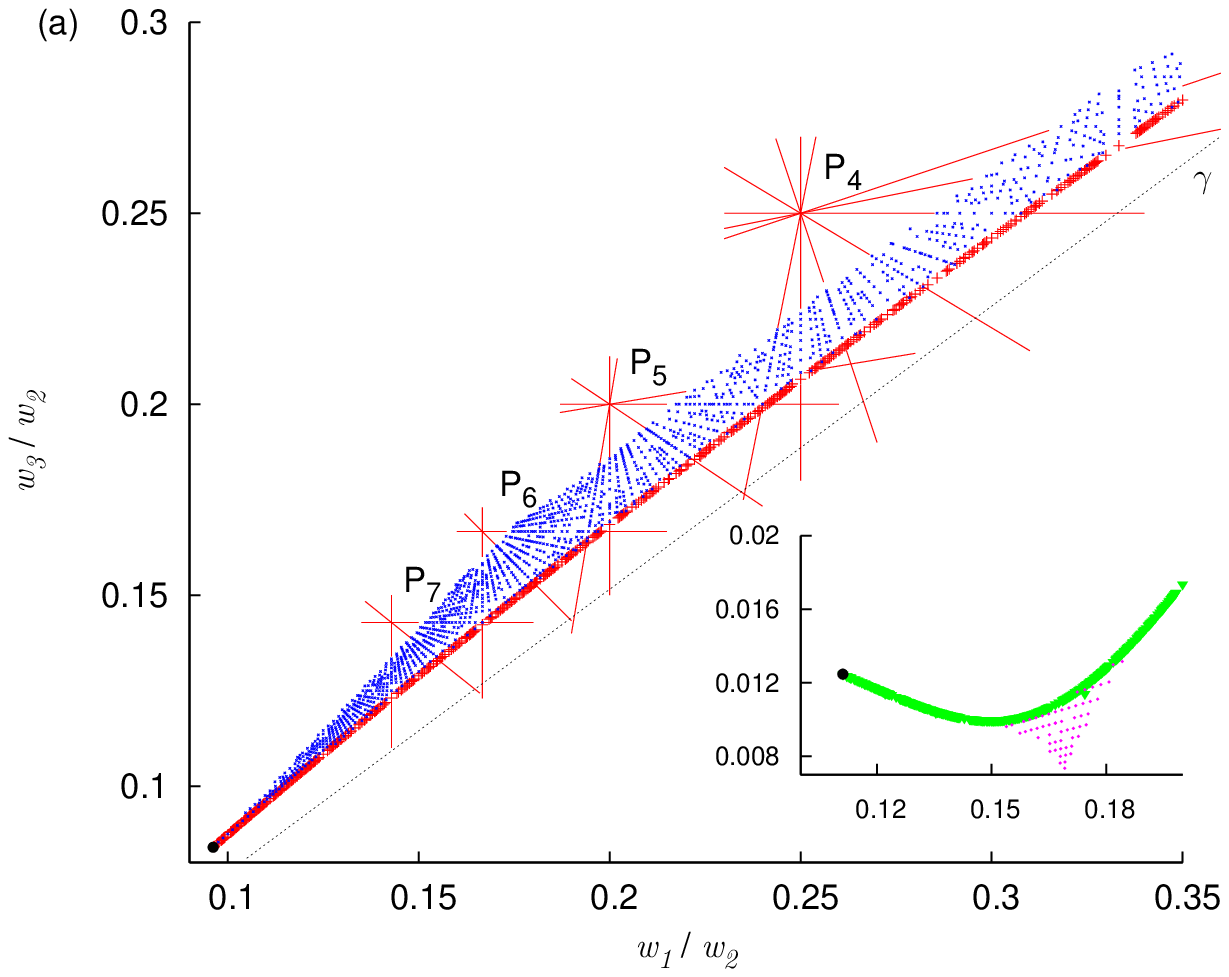}
  \includegraphics[width=.93\columnwidth]{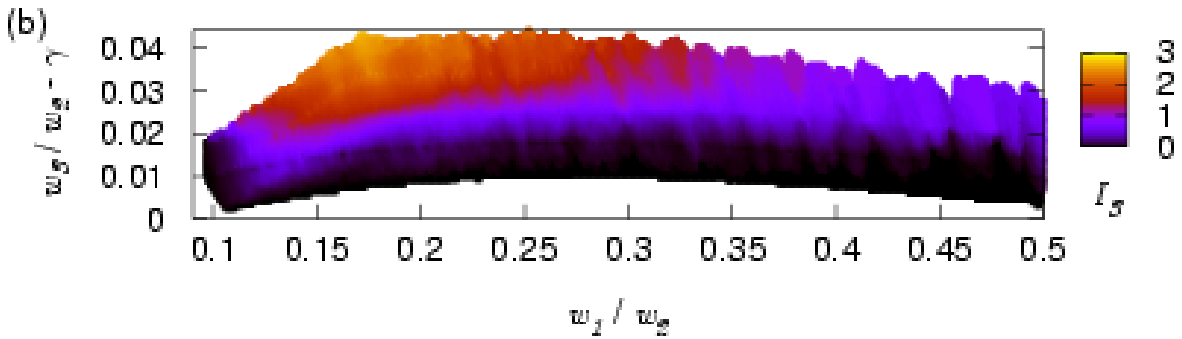}
  \caption{(color online).
  (a) Winding ratios for POs from the $T^\text{p}$ families at
  $E=-1.4\,\text{a.u.}$,
  $F=0.5\,\text{a.u.}$ with the same symbols as in Fig.~\ref{fig:TS}. 
  The most prominent resonance lines \cite{FMA_1,Resonance_1,Arnold_web}
  are indicated. The
  inset shows POs from the $T^\text{n}$ families.
  (b) Action variable $I_3$ for the $T_3^\text{p}$ family.
  For clarity the dotted line $\gamma$ from (a)
  has been subtracted on the vertical axis.}
\label{fig5}
\end{figure}

Fig.~\ref{fig5}(a) shows the winding ratios of 2- and 3-torus
POs. For the 2-torus POs, the missing ratio $w_3/w_2$ is replaced
with the stability angle that describes the dynamics transverse to the
original torus. The stability angles of the 2-torus POs limit the winding
ratios of the 3-torus POs. As this limit is approached, the action variable
$I_3$ of the 3-torus POs tends to zero, as shown in
Fig.~\ref{fig5}(b), whereas $I_1$ and $I_2$ converge toward the
values obtained from the 2-torus POs. These three observations are entirely
analogous to the three criteria laid out above that characterize the
relation of a family of 2-torus POs to its organizing FPOs.
We can therefore conclude that we are here presented with a
higher-dimensional generalization of that situation:  The $T_2$ families
serve as organizing centers for the 3-torus POs, and they impose a
distinguished coordinate system onto the $T_3$ in which $w_3$ and the action
$I_3$ identify the direction transverse to the $T_2$.

In summary, we have demonstrated how periodic orbits in a non-integrable
multidimensional Hamiltonian system can be used to reconstruct a hierarchy of
phase space structures that is organized by a stable PO at its center. The POs
on each level of this hierarchy serve as organizing centers for the
POs of the next-higher level. We have derived three general criteria to
diagnose this relation between POs. They are independent of the underlying
dynamical system and can therefore be expected to be of wide applicability. At
the same time, the calculation of the associated action variables provides the
key prerequisite for an Einstein-Brillouin-Keller torus quantization that the
crossed-fields hydrogen atom has so far resisted.  It thereby paves the way to
an immediate, and important, semiclassical application of the purely classical
results obtained here. These results will be presented in a forthcoming
publication.

\begin{acknowledgments}
We thank C.\ Chandre, \`A.\ Jorba, J.\ D.\ Meiss, and G.\ Wunner for
helpful comments and remarks.
This work was supported by the Deutsche Forschungsgemeinschaft,
Deutscher Akademischer Austauschdienst, National Science Foundation,
and the Alexander von Humboldt-Foundation.
\end{acknowledgments}

\end{document}